\documentclass[aps,prx,twocolumn,floatfix,superscriptaddress,amsmath,amssymb]{revtex4-2}

\usepackage{booktabs}
\usepackage{graphicx}
\usepackage{dcolumn}
\usepackage{bm}

\usepackage[pdftex]{color}
\usepackage[dvipsnames]{xcolor}

\usepackage[colorlinks,bookmarks=false,citecolor=Blue,linkcolor=BrickRed,urlcolor=Blue]{hyperref}
\usepackage[mathlines]{lineno}
\usepackage{braket}
\usepackage{bbm}
\usepackage{verbatim}
\usepackage{comment}
\usepackage{float} 
\usepackage{siunitx}
\usepackage{graphicx}

\usepackage{xr-hyper}
\usepackage{xcolor}

\begin{document}
\preprint{APS/123-QED}
\title{Disentangling Tensor Network States with Deep Neural Networks}

\author{Chaohui Fan} \thanks{These authors contributed equally to this work.}

\affiliation{Beijing National Laboratory for Condensed Matter Physics, Institute of Physics, Chinese Academy of Sciences, Beijing 100190, China}
\affiliation{School of Physical Sciences, University of Chinese Academy of Sciences, Beijing 100049, China}
\affiliation{ByteDance Seed, China.}

\author{Bo Zhan} \thanks{These authors contributed equally to this work.}
\affiliation{Beijing National Laboratory for Condensed Matter Physics, Institute of Physics, Chinese Academy of Sciences, Beijing 100190, China}
\affiliation{School of Physical Sciences, University of Chinese Academy of Sciences, Beijing 100049, China}
\affiliation{ByteDance Seed, China.}

\author{Yuntian Gu}  \thanks{These authors contributed equally to this work.}
\affiliation{ByteDance Seed, China.}
\affiliation{State Key Laboratory of General Artificial Intelligence, School of  Intelligence Science and Technology, Peking University.}

\author{Tong Liu}
\affiliation{Beijing National Laboratory for Condensed Matter Physics, Institute of Physics, Chinese Academy of Sciences, Beijing 100190, China}
\affiliation{School of Physical Sciences, University of Chinese Academy of Sciences, Beijing 100049, China}

\author{Yantao Wu} \thanks{yantaow@iphy.ac.cn}
\affiliation{Beijing National Laboratory for Condensed Matter Physics, Institute of Physics, Chinese Academy of Sciences, Beijing 100190, China}

\author{Mingpu Qin} \thanks{qinmingpu@sjtu.edu.cn}
\affiliation{Key Laboratory of Artificial Structures and Quantum Control (Ministry of Education),  School of Physics and Astronomy, Shanghai Jiao Tong University, Shanghai 200240, China}

\author{Dingshun Lv}\thanks{ywlds@163.com}
\affiliation{ByteDance Seed, China.}

\author{Tao Xiang}\thanks{txiang@iphy.ac.cn}
\affiliation{Beijing National Laboratory for Condensed Matter Physics, Institute of Physics, Chinese Academy of Sciences, Beijing 100190, China}
\affiliation{School of Physical Sciences, University of Chinese Academy of Sciences, Beijing 100049, China}

\date{\today}

\begin{abstract}

 We introduce Neural Tensor Network States ($\nu$TNS), a variational many-body wave-function ansatz that integrates deep neural networks with tensor-network architectures. In the $\nu$TNS framework, a neural network serves as a disentangler of the wave-function, transforming the physical degrees of freedom into renormalized variables with much less entanglement. The renormalized state is then efficiently encoded by a back-flow tensor network. This construction yields a compact yet highly expressive representation of strongly correlated quantum states. Using convolutional neural networks combined with matrix product states as a concrete implementation, we obtain state-of-the-art variational energies for the spin-$1/2$ $J_1$–$J_2$ Heisenberg model on the square lattice at the highly frustrated point $J_2/J_1=0.5$, for systems up to $20\times 20$ with periodic boundary conditions. Finite-size scaling of spin, dimer, and plaquette correlations exhibits power-law decay without magnetic or valence-bond long-range order, consistent with a gapless quantum spin-liquid ground state at that point.
 This $\nu$TNS framework is flexible and naturally extensible to other neural and tensor-network structures, offering a general platform for investigating strongly correlated quantum many-body systems.
 
\end{abstract}

\maketitle

{\em Introduction---} 
 Understanding strongly correlated quantum many-body systems remains one of the central challenges of modern physics, underlying phenomena ranging from quantum spin liquids to unconventional superconductivity. Their theoretical description is notoriously difficult, making controlled numerical simulations indispensable. Among them, tensor-network states~\cite{PhysRevLett.69.2863, 1995Ostlund, SCHOLLWOCK201196, RevModPhys.93.045003, xiang2023density} have emerged as a powerful framework for tackling strongly correlated systems, providing key insights into a wide range of problems in condensed matter physics~\cite{Dagotto1994, LeBlanc2015, 2022Chi}.

 In one dimension (1D), this success is exemplified by the density-matrix renormalization group (DMRG)~\cite{PhysRevLett.69.2863} and matrix product states (MPS)~\cite{1995Ostlund}, which together provide an optimal description of low-entanglement ground states~\cite{SCHOLLWOCK201196, RevModPhys.93.045003, xiang2023density}. In two dimensions, however, the situation is far less satisfactory. Projected entangled-pair states (PEPS)~\cite{verstraete2004renormalizationalgorithmsquantummanybody, 2008Jiang, PhysRevX.4.011025} offer a natural higher-dimensional generalization, but their computational cost rises steeply with bond dimension. It is therefore highly desirable to extend to two or higher dimensions the favorable efficiency and variational controllability of simple tensor-network states, such as MPS or tree tensor network states~\cite{Shi2006,Li2012,zhan2026}. The central obstacle is entanglement, since a highly entangled quantum state generally cannot be encoded efficiently in such a tensor network unless its entanglement is first reduced by a global disentangler.

 Neural networks provide a natural route towards this goal. Unlike tensor networks, neural networks can implement highly nonlinear, nonlocal maps in configuration space and have already demonstrated striking expressive power in quantum many-body variational wave functions~\cite{carleoSolvingQuantumManybody2017, nomuraDiracTypeNodalSpin2021a, liangHybridConvolutionalNeural2021,rothHighaccuracyVariationalMonte2023b,chenEmpoweringDeepNeural2024,wangVariationalOptimizationAmplitude2024,chenNeuralNetworkAugmentedPfaffian2025,6ccd-wzhz, geierSelfattentionNeuralNetwork2025, guSolvingHubbardModel2025}. This makes them ideally suited to reorganize complex many-body correlations before tensor-network compression. From this perspective, neural networks and tensor networks are not competing ansatzes, but can be integrated as complementary components of a single variational strategy.

 Here we introduce $\nu$TNS, a hybrid architecture that combines the global representational power of neural networks with the controlled low-rank structure of tensor networks. The key idea is to separate the tasks of disentangling and compression. A neural network first acts as a global, generally nonunitary disentangler, mapping the physical configuration to a set of renormalized variables. The residual correlations are then encoded by a tensor network, yielding a compact, efficient, and systematically improvable representation of the many-body wave function. Although previous works have explored different combinations of neural networks and tensor networks~\cite{du2025neuralizedfermionictensornetworks, liangHybridConvolutionalNeural2021, PhysRevResearch.5.L032001, chen2024antnbridgingautoregressiveneural}, this ansatz establishes a genuinely synergistic framework in which the neural network performs global disentangling and the tensor network encodes the residual correlations.
 
 As a representative example, we implement this ansatz using convolutional neural networks (CNNs) combined with MPS (CNN-MPS) and demonstrate the potential of this approach on the spin-1/2 $J_1–J_2$ Heisenberg model on the square lattice at the highly frustrated point $ J_2/J_1 = 0.5$, whose ground state has remained under debate for more than two decades. Competing proposals include valence-bond-solid order, antiferromagnetism, and gapless spin liquids, and no consensus has emerged ~\cite{gongPlaquetteOrderedPhase2014,haghshenas1SymmetricInfinite2018,liuGaplessQuantumSpin2022a,liuGaplessSpinLiquid2018,nomuraDiracTypeNodalSpin2021a,qianAbsenceSpinLiquid2024,schulzMagneticOrderDisorder1996,wangCriticalLevelCrossings2018, SM}. Using the CNN-MPS ansatz, we obtain the lowest variational energies reported to date for systems up to $20 \times 20$. Moreover, finite-size scaling of spin, dimer, and plaquette correlations exhibits power-law decay without any sign of long-range order, consistent with a gapless quantum spin-liquid ground state at $J_2/J_1 = 0.5$.

\vspace{2mm}

\begin{figure}
    \centering
    \includegraphics[width=0.8\linewidth]{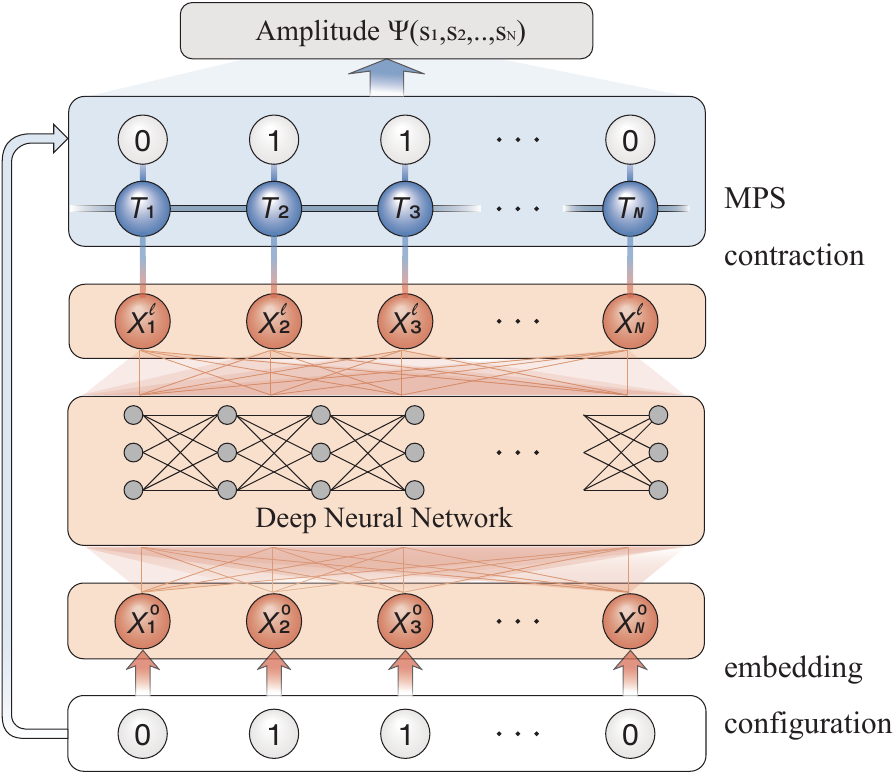}
    \caption{Graphical illustration of the $\nu$TNS ansatz with MPS as a concrete realization of TNS. The physical configuration is first mapped into feature space by a linear embedding, yielding the initial feature vector $X^{0}$. This vector is then processed through $\ell$ neural-network layers to produce the renormalized feature vector $X^{\ell}$, which is finally contracted with a backflow MPS to generate the $\nu$MPS wave function.}
    \label{fig:nu_mps}
\end{figure}

{\em $\nu$TNS ansatz---}
 The key idea of $\nu$TNS is to separate global correlation extraction from entanglement compression. Figure~\ref{fig:nu_mps} schematically illustrates the ansatz, with an MPS serving as a concrete realization of the tensor network layer. Starting from a physical configuration on a lattice of $N$ sites, we first embed the local variables into a trainable feature vector $X^{(0)}\in\mathbb{R}^{N\times h}$, where $h$ is the hidden dimension. This embedding consists of a learned linear map augmented with positional encoding, which lifts discrete variables into a continuous space and incorporates lattice geometry. The feature vector $X^{(0)}$ is then processed by $\ell$ neural-network layers, for example $\ell$ convolutional layers in CNN, to produce $X^{(\ell)}$. Although $X^{(\ell)}$ is formally an $N$-site product state in feature space, it contains highly nonlocal information, since each site feature has aggregated correlations from distant regions through the neural network. The neural network here acts as a global disentangler, in contrast to the local disentangling transformations considered in Refs.~\cite{Qian_2023,PhysRevLett.133.190402,2025arXiv250508635H}.

 To convert $X^{(\ell)}$ into a wave-function amplitude, we use an MPS as the final layer. For each site $i$, we introduce a rank-4 tensor $T_i\in\mathbb{R}^{h\times D\times D\times d}$, where $d$ is the physical dimension and $D$ is the MPS bond dimension. Contracting $T_i$ with the local feature vector and the initial configuration yields the corresponding MPS tensor $A_i$, from which the wave-function amplitude is obtained as
\begin{equation}
\Psi(s_1,\ldots,s_N)=\mathrm{Tr}\prod_{i=1}^N A_i[s_1,\ldots,s_N].
\end{equation}
 Because the local feature vector depends on the full physical configuration $(s_1,\ldots,s_N)$, the resulting tensor $A_i$ is also configuration dependent, unlike in ordinary MPS where the local tensor depends on the on-site variable $s_i$ only. This construction therefore realizes a backflow MPS~\cite{PhysRev.102.1189,PhysRevB.78.041101,luoBackflowTransformationsNeural2019} for each input configuration. For a fixed neural network, the bond dimension $D$ serves as a control parameter for the representational capacity of $\nu$TNS. More details about the ansatz can be found in the Supplementary Materials.

All variational parameters are optimized using the MARCH algorithm~\cite{guSolvingHubbardModel2025}, a stabilized variant of stochastic reconfiguration. In addition, lattice symmetries, such as the $C_{4v}$ point group, can be imposed to improve the quality of the variational state and enhance the stability of the optimization.

\vspace{2mm}

{\em Result and discussion---}
 We benchmark the $\nu$TNS ansatz on the spin-$1/2$ $J_1-J_2$ Heisenberg model on the $N=L\times L$ square lattice,
\begin{equation}
  H=J_1\sum_{\langle i,j\rangle} \vec{S}_i\!\cdot\! \vec{S}_j +J_2\sum_{\langle\langle i,j\rangle\rangle}\vec{S}_i\!\cdot\!\vec{S}_j,
\end{equation}
 where $\vec{S}_i$ denotes the spin-$1/2$ operator at site $i$, and $\langle i,j\rangle$ and $\langle\langle i,j\rangle\rangle$ represent nearest- and next-nearest-neighbor pairs, respectively. We impose periodic boundary conditions and set $J_1=1$ as the unit of energy. In the following we focus on the point $J_2/J_1=0.5$.

\begin{figure}
    \centering
    \includegraphics[width=1\linewidth]{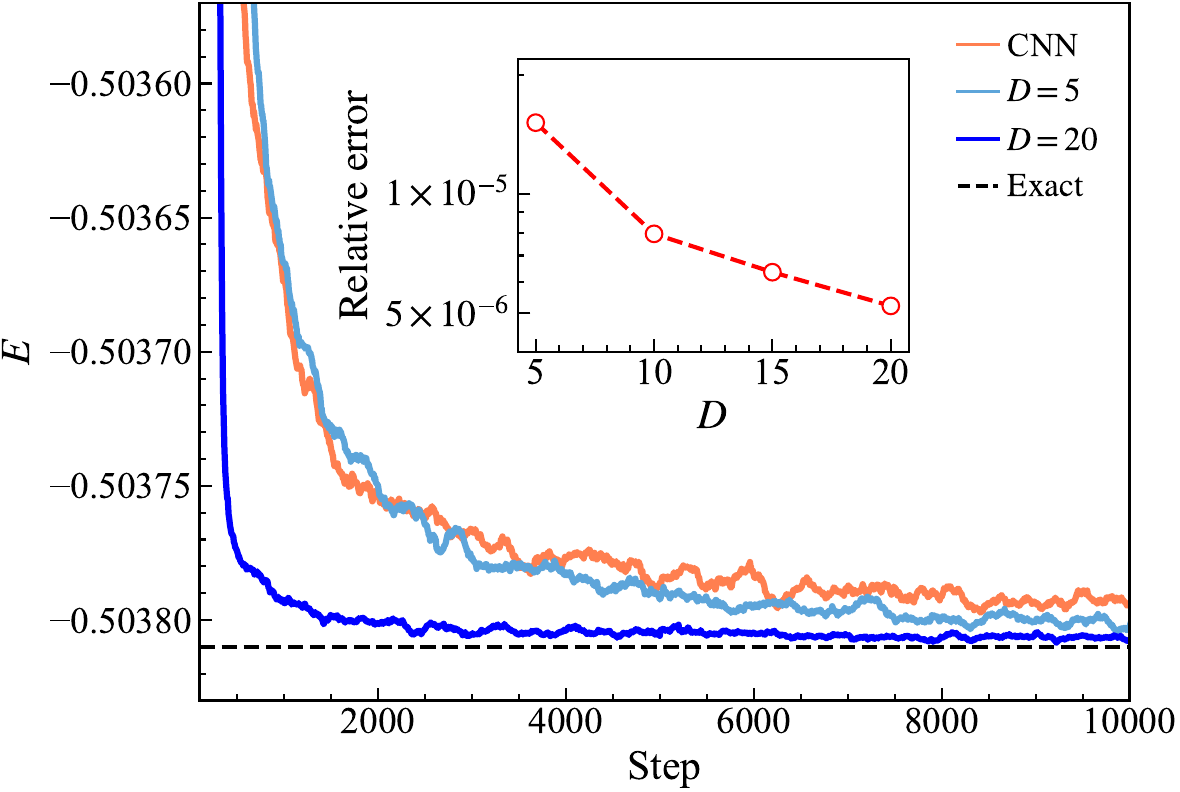}
    \caption{Convergence of the ground state energy with the optimization step for the $6\times 6$ $J_1-J_2$ Heisenberg model with periodic boundary conditions at $J_2/J_1=0.5$, obtained using the pure CNN and the CNN-MPS ansatz with $D=5$ and 20. The dotted horizontal line marks the exact result~\cite{schulzMagneticOrderDisorder1996}. Inset: relative energy error versus $D$.}
    \label{fig:benchmark_Conv}
\end{figure}

 To characterize the spin, dimer, and plaquette orders, we calculate the correlation functions in the ground state,
\begin{eqnarray}
C_s(i,j) &=& \langle \vec{S}_i \!\cdot\! \vec{S}_j \rangle, \nonumber \\
C_d(i,j) &= &\langle D_i D_j \rangle 
           - \langle D_i \rangle \langle D_j \rangle, \\
C_p(i,j) &= & \langle P_i P_j \rangle 
           - \langle P_i \rangle \langle P_j \rangle , \nonumber
\end{eqnarray}
 where $D_i = \vec{S}_i \!\cdot\! \vec{S}_{i+\hat{x}}$, $P_i = (P_{\square_i} + P_{\square_i}^{-1})/2$, and $P_{\square_i}$ is the cyclic permutation operator acting on the four spins around plaquette~$i$. In $C_d(i,j)$ we choose the bond orientation along the $x$ direction; the results for the $y$ direction are identical owing to the preserved $C_4$ lattice rotation symmetry. The corresponding structure factors are defined as
\begin{equation}
  S_\gamma(q) =\frac{1}{N}\sum_{i,j}C_\gamma(i, j)\, e^{i q \cdot (r_i-r_j)}, \quad (\gamma=s,d,p)
\end{equation} 
 which provide quantitative diagnostics of the dominant ordering channels.

 We first assess the accuracy of CNN-MPS on small clusters where reliable benchmarks are available. For the $6\times 6$ lattice, exact diagonalization gives the ground-state energy $-0.503810$ and the spin structure factor $S_s(\pi,\pi)=3.50968$~\cite{schulzMagneticOrderDisorder1996}. For the pure CNN wave function with $(h,\ell) = (32,20)$, which corresponds to setting $D=1$ in CNN-MPS, we obtain an energy of $-0.5037957(4)$ [Fig.~\ref{fig:benchmark_Conv}], with a relative error of $3 \times 10^{-5}$. Increasing the MPS bond dimension yields systematic improvement. At $D=20$, the energy reaches $-0.50380737(5)$, reducing the relative error to $5\times 10^{-6}$. Achieving comparable accuracy with a pure MPS requiring a bond dimension of order $10^4$~\cite{gongPlaquetteOrderedPhase2014}. CNN-MPS also reproduces the spin structure factor, $S_s(\pi,\pi)=3.50967(2)$, in good agreement with the exact value. These results provide direct evidence for the central idea of $\nu$TNS that CNN efficiently extracts and compresses the dominant long-range correlations, leaving only a much weaker residual entanglement to be captured by an MPS with modest bond dimension.

\begin{figure}
    \centering
    \includegraphics[width=0.9\linewidth]{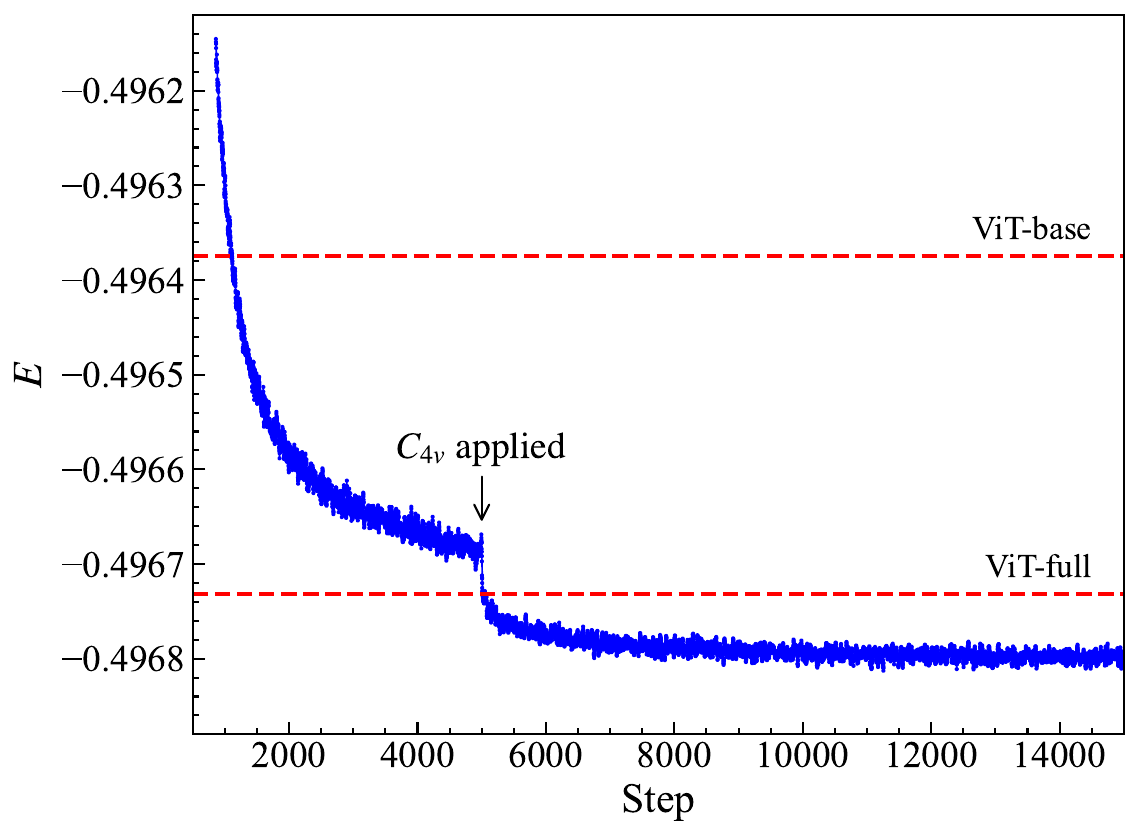}
    \caption{Optimization of the ground-state energy for the $L=20$ $J_1-J_2$ Heisenberg model with the CNN-MPS ansatz. The $C_{4v}$ lattice symmetry is enforced after the first $5{,}000$ steps. Also shown for comparison are the ViT results reported in Ref.~\cite{viteritti2026approachingthermodynamiclimitneuralnetwork}, where ViT-base and ViT-full refer to calculations without and with full point-group symmetry, respectively.}
    \label{fig:3}
\end{figure}

 For the $10\times 10$ lattice, where a large body of variational results is available in the literature (Table~\ref{tab:energy_comparison}), CNN-MPS again achieves state-of-the-art accuracy. With the Marshall sign transformation~\cite{Antiferromagnetism1955}, the best energy obtained by CNN-MPS is $-0.4976939(2)$ for $(h,D,\ell)=(32,20,20)$, which is lower than the best previously reported result~\cite{chenEmpoweringDeepNeural2024}.

\begin{table}[b]
    \centering
    \caption{Ground-state energy per site $E$ for the square-lattice $J_1-J_2$ Heisenberg model at $J_2/J_1=0.5$ with periodic boundary conditions.}
    \label{tab:energy_comparison}
    \begin{tabular*}{\columnwidth}{@{\extracolsep{\fill}} c l c r c c}
        \hline\hline
        L & $\quad\quad E$ & Wave function &Year  &Ref. \\
        \hline
        10      & -0.49516(1)    & CNN  &2019  &~\cite{PhysRevB.100.125124}  \\  
                & -0.495502(1)   & PEPS+CNN &2021 &~\cite{liangHybridConvolutionalNeural2021} \\
                & -0.495530      & DMRG     &2014 &~\cite{gongPlaquetteOrderedPhase2014}\\
                & -0.495627(6)   & aCNN &2024 &~\cite{wangVariationalOptimizationAmplitude2024} \\
                & -0.49575(3)    & RBM  &2019  &~\cite{PhysRevB.100.125131} \\
                & -0.49586(4)    & CNN  &2023  &~\cite{PhysRevB.107.195115} \\
                & -0.4968(4)     & RBM  &2022  &~\cite{chen2022systematicimprovementneuralnetwork} \\
                & -0.49717       & CNN  &2022  &~\cite{Li_2022} \\
                & -0.497437(7)   & GCNN     &2023 &~\cite{rothHighaccuracyVariationalMonte2023b} \\
                & -0.49747       & CNN  &2023 &~\cite{liangDeepLearningRepresentations2023}      \\
                & -0.49755(1)    & VMC  &2013  &~\cite{PhysRevB.88.060402}  \\
                & -0.497629(1)   & RBM+PP &2021 &~\cite{nomuraDiracTypeNodalSpin2021a} \\
                & -0.497634(1)   & ViT      &2024 &~\cite{rendeSimpleLinearAlgebra2024} \\
                & -0.4976764(7)  & CTWF &2025 &~\cite{chen2025convolutionaltransformerwavefunctions}\\
                & -0.4976921(4)  & CNN  &2024  &~\cite{chenEmpoweringDeepNeural2024} \\
                & \hspace{-0.1em}-\scalebox{0.9}[1]{\textbf{0.4976923}}(2) & T-MPS   &    & this work \\
                & \hspace{-0.1em}-\scalebox{0.9}[1]{\textbf{0.4976939}}(2) & CNN-MPS &    & this work\\
        \hline
        16      & -0.49626      & CNN &2022 &~\cite{Li_2022} \\
                & -0.49659      & CNN &2023 &~\cite{liangDeepLearningRepresentations2023} \\
                & -0.4967163(8) & CNN &2024 &~\cite{chenEmpoweringDeepNeural2024} \\
                & \hspace{-0.1em}-\scalebox{0.9}[1]{\textbf{0.496786}}(1)  & T-MPS &  & this work\\
                & \hspace{-0.1em}-\scalebox{0.9}[1]{\textbf{0.4969140}}(5)  & CNN-MPS&  & this work\\
        \hline
        20      & -0.496732(1)  &ViT    &2026   &~\cite{viteritti2026approachingthermodynamiclimitneuralnetwork}\\
                & \hspace{-0.1em}-\scalebox{0.9}[1]{\textbf{0.4967987}}(6) & CNN-MPS&       & this work\\

        \hline\hline
    \end{tabular*}
\end{table}

 The advantage of CNN-MPS becomes even more evident on larger lattices. Using the same parameters, $(h,D, \ell)=(32,15,20)$, for all system sizes, the total number of variational parameters remains nearly constant, apart from the small position-encoding contribution, yet the achieved accuracy continues to improve over the best results in the literature. For $L=16$, CNN-MPS yields the lowest variational energy, $-0.4969140(5)$, compared with the previously best reported value $-0.4967163(8)$~\cite{chenEmpoweringDeepNeural2024}. For the largest system considered, $L=20$, we compare with the recent neural-network results reported in  Ref.~\cite{viteritti2026approachingthermodynamiclimitneuralnetwork}, where energies with and without symmetry projection were computed. After $5{,}000$ optimization steps without any imposed symmetry and a further $10{,}000$ steps with $C_{4v}$ symmetry enforced, we obtain $-0.4966857(7)$ and $-0.4967987(6)$, respectively, both significantly below the corresponding values reported in Ref.~\cite{viteritti2026approachingthermodynamiclimitneuralnetwork}, as shown in Fig.~\ref{fig:3}. These results show that the $\nu$TNS construction is not only highly accurate, but also scalable as the lattice size increases. The overall computational cost scales as $O(N^2)$, making the method suitable for large-scale simulations.

\begin{figure}
    \centering
    \includegraphics[width=0.9\linewidth]{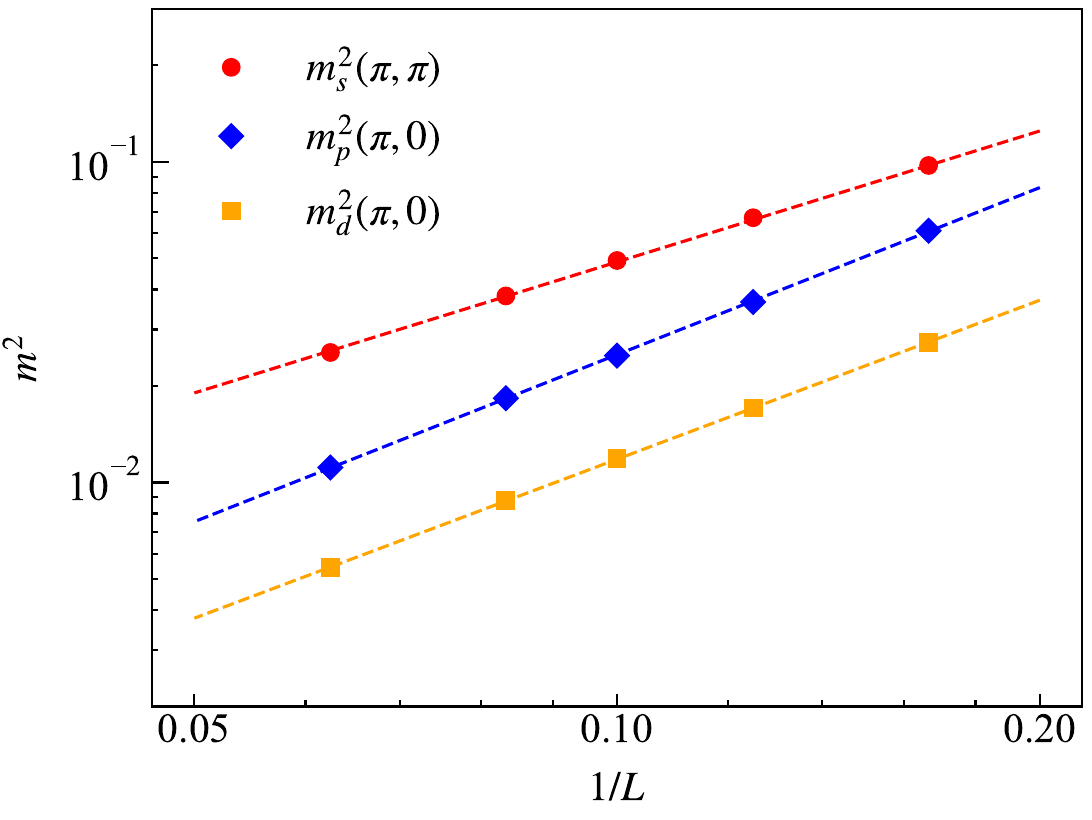}
    \caption{Log-log plot of the finite-size scaling of the order parameters \(m_s^2(\pi,\pi)\), \(m_p^2(\pi,0)\), and \(m_d^2(\pi,0)\) for the \(J_1-J_2\) Heisenberg model at \(J_2/J_1=0.5\), obtained using the CNN-MPS ansatz with $(h, D, \ell)=(32, 15, 20)$.
    }
    \label{fig:4}
\end{figure}
 
 Having established the high quantitative accuracy of the ansatz, we next turn to the nature of the ground state at $J_2/J_1=0.5$. To distinguish among the possible competing orders, we analyze the three structure factors. The plaquette structure factor $S_p(\mathbf{q})$ displays maxima at $(\pi,0)$ and $(0,\pi)$, but it shows no appreciable peak at $(\pi,\pi)$. Since a pronounced $(\pi,\pi)$ peak is the characteristic fingerprint of a plaquette valence-bond phase, our results rule out the presence of this order in the ground state. By contrast, the dimer structure factor $S_d(\mathbf{q})$ develops its dominant peak at $(\pi,0)$, indicating that the leading valence-bond-solid fluctuation channel is columnar rather than plaquette-like.

 A more quantitative picture is obtained from the finite-size scaling of the order parameters extracted from the structure factors, $m_\gamma^2(\mathbf{q})= S_\gamma(\mathbf{q}) / N $,  evaluated at the dominant ordering wave vectors. Specifically, the leading peaks occur at $(\pi,\pi)$ for $\gamma=s$, and at $(\pi,0)$ for both $\gamma=(d, p)$. As shown in Fig.~\ref{fig:4}, all three order parameters decay approximately linearly with $1/L$ on log-log plots, corresponding to power-law scaling in ordinary coordinates. This behavior indicates the absence of true long-range order in these three channels in the thermodynamic limit. The slow power-law decay reveals substantial critical correlations, consistent with a gapless quantum spin liquid rather than a conventionally ordered antiferromagnet or valence-bond-solid phase. Thus the $\nu$TNS results support a ground state at $J_2/J_1=0.5$ with strong antiferromagnetic and valence-bond correlations but no long-range order.
 
 We have also explored an alternative realization of $\nu$TNS in which the CNN is replaced by a transformer-based neural network, which we denote as T-MPS. The corresponding results are summarized in Table~\ref{tab:energy_comparison}. This T-MPS ansatz consistently improves upon the corresponding neural-network ansatz with the MPS back-end, further confirming the usefulness of the $\nu$TNS strategy. However, for the present problem, T-MPS is less accurate and computationally more expensive than CNN-MPS. This comparison suggests that, for the frustrated $J_1-J_2$ Heisenberg model considered here, convolutional architectures are more effective than transformer-based ones at disentangling the relevant entanglement. 

\vspace{2mm}

{\em Conclusion and perspectives.---}
 In conclusion, we have introduced $\nu$TNS as a flexible and general variational framework that unifies neural-network and tensor-network representations for quantum many-body states. By embedding physical configurations into a feature space, processing them through a deep neural network, and contracting the resulting renormalized features with a backflow-enhanced tensor network, $\nu$TNS achieves a hierarchical description that combines expressive power with efficient entanglement control. In its CNN-MPS realization, the method yields state-of-the-art results for the frustrated $J_1-J_2$ Heisenberg model at $J_2/J_1=0.5$ on periodic lattices up to $20\times 20$. The finite-size scaling analysis of spin, dimer, and plaquette correlations shows power-law decay in all channels and no evidence for the corresponding long-range orders, consistent with the scenario of a gapless quantum spin-liquid ground state at this highly frustrated point.
 
 Beyond this specific application, the main strength of the $\nu$TNS framework lies in its versatility. The neural network can be replaced by alternative architectures better suited to different correlation patterns, while the tensor-network backbone can be upgraded from MPS to tree tensor networks, PEPS, or other structures adapted to the geometry and entanglement properties of the target system. This flexibility opens a broad range of future directions. Particularly promising are applications to fermionic systems, where the combination of neural-network-based correlation extraction and tensor-network-based entanglement compression may provide a powerful route to treating sign structures and long-range quantum correlations on equal footing. It will also be important to extend the framework to larger system sizes, excited states, dynamical properties, and finite-temperature problems, as well as to develop more efficient optimization strategies and symmetry-preserving implementations. With continued advances in neural network architectures, tensor network algorithms, and large-scale optimization techniques, $\nu$TNS offers a practical and scalable pathway for investigating strongly correlated quantum matter and identifying exotic quantum phases beyond the reach of conventional methods.

{\em Acknowledgments:} 
We acknowledge the support from the National Natural Science Foundation of China (Grant Nos. 12488201, 12522406, and 12274290), the National Key Research and Development Program of China (2021ZD0301800, 2022YFA1405400), and the Innovation Program for Quantum Science and Technology (2021ZD0301902).
Y.W. is supported by a start-up grant from IOP-CAS.

\bibliography{refs}

\end{document}


\title{
Disentangling Tensor Network States with Deep Neural Networks \\ 
Supplementary Materials }

\author{Chaohui Fan} \thanks{These authors contributed equally to this work.}
\affiliation{Beijing National Laboratory for Condensed Matter Physics, Institute of Physics, Chinese Academy of Sciences, Beijing 100190, China}
\affiliation{School of Physical Sciences, University of Chinese Academy of Sciences, Beijing 100049, China}
\affiliation{ByteDance Seed, China.}

\author{Bo Zhan} \thanks{These authors contributed equally to this work.}
\affiliation{Beijing National Laboratory for Condensed Matter Physics, Institute of Physics, Chinese Academy of Sciences, Beijing 100190, China}
\affiliation{School of Physical Sciences, University of Chinese Academy of Sciences, Beijing 100049, China}
\affiliation{ByteDance Seed, China.}

\author{Yuntian Gu}  \thanks{These authors contributed equally to this work.}
\affiliation{ByteDance Seed, China.}
\affiliation{State Key Laboratory of General Artificial Intelligence, School of  Intelligence Science and Technology, Peking University.}

\author{Tong Liu}
\affiliation{Beijing National Laboratory for Condensed Matter Physics, Institute of Physics, Chinese Academy of Sciences, Beijing 100190, China}
\affiliation{School of Physical Sciences, University of Chinese Academy of Sciences, Beijing 100049, China}

\author{Yantao Wu} \thanks{yantaow@iphy.ac.cn}
\affiliation{Beijing National Laboratory for Condensed Matter Physics, Institute of Physics, Chinese Academy of Sciences, Beijing 100190, China}

\author{Mingpu Qin} \thanks{qinmingpu@sjtu.edu.cn}
\affiliation{Key Laboratory of Artificial Structures and Quantum Control (Ministry of Education),  School of Physics and Astronomy, Shanghai Jiao Tong University, Shanghai 200240, China}

\author{Dingshun Lv}\thanks{ywlds@163.com}
\affiliation{ByteDance Seed, China.}

\author{Tao Xiang}\thanks{txiang@iphy.ac.cn}
\affiliation{Beijing National Laboratory for Condensed Matter Physics, Institute of Physics, Chinese Academy of Sciences, Beijing 100190, China}
\affiliation{School of Physical Sciences, University of Chinese Academy of Sciences, Beijing 100049, China}

\maketitle

\setcounter{figure}{0}
\setcounter{table}{0}
\setcounter{equation}{0}
\setcounter{section}{0}
\renewcommand{\thesection}{S\arabic{section}}
\renewcommand{\thefigure}{S\arabic{figure}}
\renewcommand{\thetable}{S\arabic{table}}
\renewcommand{\theequation}{S\arabic{equation}}

\makeatletter
\renewcommand{\@biblabel}[1]{[S#1]}
\makeatother

\section{phase diagram of the square Heisenberg $J_1-J_2$ model}

 Quantum spin liquids (QSLs) are paradigmatic strongly correlated quantum many-body states, distinguished by long-range quantum entanglement, fractionalized excitations, and nontrivial topological order. Establishing their existence in realistic lattice models remains notoriously difficult because of the close competition from nearly degenerate phases, such as valence-bond solids (VBSs) and antiferromagnetic (AFM) states with weak order. 

\begin{figure}[b]
    \centering
    \includegraphics[width=1\linewidth]{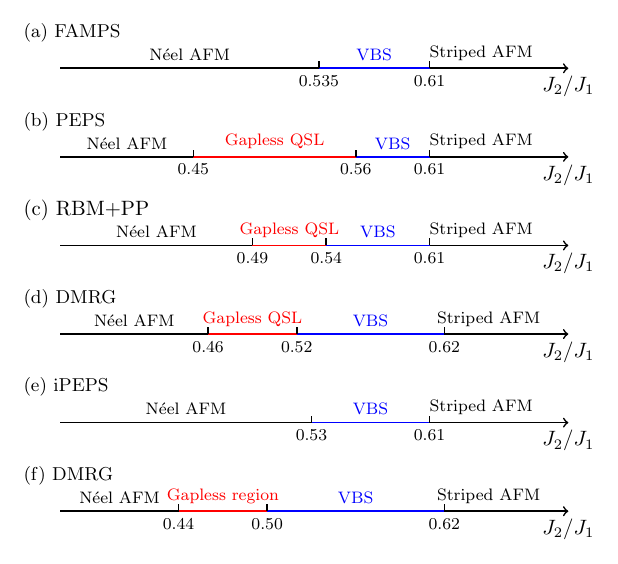}
    \caption{Phase diagrams for square Heisenberg $J_1-J_2$ model obtained by different methods: (a) from Ref.~\cite{qianAbsenceSpinLiquid2024}, (b) from Ref.~\cite{liuGaplessQuantumSpin2022a}, (c) from Ref.~\cite{nomuraDiracTypeNodalSpin2021a}, (d) from Ref.~\cite{wangCriticalLevelCrossings2018}, (e) from Ref.~\cite{morawetzU1SymmetricRecurrent2021}, and (f) from Ref.~\cite{gongPlaquetteOrderedPhase2014}. Controversy exists in the vicinity of the highly frustrated point $J_2/J_1=0.5$.}
    \label{fig:phase}
\end{figure}

 The spin-1/2 $J_1$–$J_2$ Heisenberg model on the square lattice provides a canonical setting for investigating this competition. As illustrated in Fig.~\ref{fig:phase}, although the precise phase boundaries vary among numerical studies, the regime $0.4 \lesssim J_2/J_1 \lesssim 0.6$ is consistently identified as strongly frustrated, with the point $J_2/J_1 = 0.5$ remaining particularly controversial. We therefore use this point as a benchmark to demonstrate the potential of our method.

 The spin, dimer, and plaquette correlations presented in the main text all exhibit power-law decay, without evidence of magnetic or valence-bond long-range order. These results are consistent with a gapless quantum-spin-liquid state at $J_2=0.5J_1$.

\section{The $\nu$TNS ansatz}

 In the $\nu$TNS ansatz, the many-body wave function is given by 
\begin{equation}
    |\Psi\rangle = \sum_{s_1, \cdots , s_N} \Psi(s_1, \cdots , s_N) |s_1, \cdots , s_N\rangle ,\label{eq:wave_func}
\end{equation}
 where $\Psi $ is parametrized by a neural network augmented tensor network state. A schematic illustration of $\nu$TNS is shown in Fig.~1 of the main text. Here we use MPS as a concrete realization of the tensor-network component to demonstrate how the ansatz works. In particular, we consider the two specific implementations discussed in the main text, namely T-MPS and CNN-MPS.

\subsection{Embedding of physical degrees}

 For $\nu$TNS, each spin-$1/2$ degree of freedom is first represented by a one-hot vector, $(1,0)^T$ or $(0,1)^T$, and embedded into an $h$-dimensional feature space as
\begin{equation}
 X_i^0 = v_i W_0 + E_i, \qquad
 W_0\in\mathbb{R}^{2\times h}; E\in\mathbb{R}^{N\times h},
\end{equation}
 where $E_i$ provides position encoding. These site features forms the vector $X^{(0)}\in\mathbb{R}^{N\times h}$ and serve as the input to the neural-network layers.

\subsection{Transformer layers}

In T-MPS, after embedding the architecture proceeds with $\ell$ composite layers, each comprising a multi-head self-attention module and a subsequent feed-forward network (FFN)~\cite{NIPS2017_3f5ee243}. Each layer is defined as
\begin{eqnarray}
    Y^{(l)}&=& X^{(l-1)}+\mathrm{Att}^{(l)}\left(\mathrm{LN}\left(X^{(l-1)}\right)\right), \\
    X^{(l)}&=& Y^{(l)}+\mathrm{FFN}^{(l)}(Y^{(l)}),
\end{eqnarray}
where LN denotes layer normalization~\cite {ba2016layernormalization} and the attention block is given by
\begin{eqnarray}
\mathrm{Att}^{(l)}(X) &=&  \sum_{m=1}^H W_{\mathrm{weight}}^{(l,m)}
X W_V^{(l,m)} (W_O^{(l,m)})^T, \\
W_{\mathrm{weight}}^{(l,m)} &= &
\mathrm{softmax}\left[ \frac{XW_Q^{(l,m)} (XW_K^{(l,m)})^{T}}{\sqrt{h/H}} \right],
\end{eqnarray}
where $W_Q^{(l,m)},W_K^{(l,m)},W_V^{(l,m)},W_O^{(l,m)}
\in\mathbb{R}^{h\times h/H}$ are variational parameters. The FFN block is
\begin{equation}\label{FFN}
\mathrm{FFN}^{(l)}(X)=\sigma(XW_F^{(l)} + b^{(l)})
\end{equation}
with activation $\sigma=\mathrm{SiLU}$. 

The output $X^{(\ell)}$ is a $N\times h$ matrix. Technically, the transformer block enables information teleportation across all pairs of sites through learnable query, key, and value projections. Physically, this mechanism can be interpreted as a global ``correlation extractor''. The attention weights $W_{\mathrm{weight}}$ dynamically determine which spin pairs correlate more strongly for a given configuration.

\subsection{Convolutional layers}

 A CNN layer is defined by a local linear transformation followed by a non-linear activation:
\begin{equation}
    X^{(l)} = \mathrm{LN}\left(\sigma \left(\mathrm{Conv}^{(l)}(X^{(l-1)})\right)+ X^{(l-1)} \right).
\end{equation}
The convolution operation applies a sliding window across the system. For example, for a local trainable $3 \times 3$ kernel with weights $W^{(l)}$ and a bias $b^{(l)}$, the transformation at spatial site $(i,j)$ is given by:
\begin{equation}
    \mathrm{Conv}^{(l)}(X^{(l-1)})_{i,j} = \sum_{m,n=-1}^{1} W_{m,n}^{(l)} X_{i+m, j+n}^{(l-1)} + b^{(l)}.
\end{equation}
Before the contraction with MPS, a FFN block, Eq.~\ref{FFN}, is applied. 

 This structure is simpler than the attention-based construction and resembles a renormalization procedure. The local spatial summation effectively acts as a block-spin decimation, allowing coarse-graining to be achieved through successive convolutional transformations.

\subsection{Output layer}

 In the output layer, we introduce a matrix product operator (MPO) with local tensors $T_i \in \mathbb{R}^{h\times D\times D\times d}$ at lattice sites $i$, which maps the final feature vector $X^{(\ell)}$ to a configuration-dependent (backflow) periodic MPS with local tensors $A_i$. Contracting this MPS yields the wave-function amplitude \eqref{eq:wave_func}. 

  To further enhance the representational power, we introduce $n$ independent MPOs and contract each of them with $X^{(\ell)}$. This construction produces $n$ independent MPS with the same bond dimension $D$, and the wave function is taken as the sum of these MPS. In this work, we set $n=2$.

\section{Variational optimization}

\subsection{The MARCH algorithm}

 We optimize the variational parameters using the MARCH algorithm~\cite{guSolvingHubbardModel2025}, which is an extension of stochastic reconfiguration (SR). In SR, the parameter update is determined by minimizing the Fubini–Study distance between the imaginary-time-evolved state $e^{-\delta\tau H}|\Psi_\theta\rangle$ and the variational state $|\Psi_{\theta+\mathrm{d}\theta}\rangle$ within the ansatz manifold.

 In MARCH, the parameter update $\mathrm{d}\theta_k$ is obtained by minimizing
 \begin{eqnarray}
  \mathrm{d}\theta_k &=& \min_{\mathrm{d}\theta’} \left[ \frac{1}{\lambda}|\mathbf{O}\mathrm{d}\theta’-\boldsymbol{\epsilon}|^2  \right. \nonumber \\ 
  && \left. +  |\mathrm{diag}(V_{k-1})^{\frac{1}{4}} (\mathrm{d}\theta’-\phi_{k-1})|^2 \right] .
  \label{eq:march}
 \end{eqnarray}
 The first term is the usual SR objective, while the second incorporates momentum and adaptive regularization, which stabilizes the optimization for large parameter spaces. Here, $\phi_k=\mu,\mathrm{d}\theta_k$ is the momentum term, and
\begin{equation}
   V_k=\beta V_{k-1}+(\mathrm{d}\theta_k-\mathrm{d}\theta_{k-1})^2
\end{equation}
 is an adaptive variance estimator that controls the effective step size. The parameters  used in this work are listed in Table~\ref{tab:opt}, unless otherwise specified.

\begin{table}[t]
\centering
\renewcommand{\arraystretch}{1.2}
\caption{Hyperparameters used in the optimization.}
\label{tab:opt}
\begin{tabular}{l|l}  
\toprule\toprule
\textbf{Config} & \textbf{Value} \\
\midrule
Optimizer & MARCH \\
Optimizer momentum & $\mu,\ \beta = 0.98,\ 0.995$ \\
Damping & $\lambda = 0.001$ \\
Imaginary time step length & $\tau=0.1$ \\
Batch size & 4096 \\
Learning rate & $lr = 0.15$ for CNN-MPS \\
              & $lr = 0.10$ for T-MPS \\
Norm constraint at time $t$ & $lr*\left(1 + \frac{\max(t-t_0)}{3000}\right)^{-1}$ \\
                & $t_0 = 4000$ for CNN-MPS \\
                & $t_0 = 9000$ for T-MPS \\
MCMC step & $2L_x \times L_y$ \\
Steps & 15000 \\
\bottomrule\bottomrule
\end{tabular}
\end{table}

 The quantities entering Eq.~\eqref{eq:march} are evaluated from Monte Carlo samples of configurations $\mathbf{s}$. The effective imaginary-time force is 
\begin{equation}
\epsilon(\mathbf{s}) = -\delta\tau\bigl(E_\mathrm{loc}(\mathbf{s})-\mathbb{E}_\mathbf{s}[E_\mathrm{loc}(\mathbf{s})]\bigr),
\end{equation}
with $E_{\mathrm{loc}}(\mathbf{s})$ the local energy. The logarithmic derivatives 
\begin{equation}
\bar{O}_k(\mathbf{s})=\frac{\partial}{\partial\theta_k}\log|\Psi_\theta(\mathbf{s})|,
\end{equation}
and their centered counterparts,
\begin{equation}
O_k(\mathbf{s})=\bar{O}_k(\mathbf{s})-\mathbb{E}_\mathbf{s}[\bar{O}_k(\mathbf{s})],
\end{equation}
define the SR operator vector 
\begin{equation}
O(\mathbf{s})=(O_1(\mathbf{s}),\cdots,O_k(\mathbf{s}),\cdots)^T.
\end{equation}
For a batch of sampled configurations ${\mathbf{s}_1,\ldots,\mathbf{s}_B}$, we define operator matrix $\mathbf{O}$ and force vector $\boldsymbol{\epsilon}$ as
\begin{eqnarray}
    \mathbf{O} &=&  
\begin{pmatrix}
  O(\mathbf{s}_1) ,  O(\mathbf{s}_2) , \cdots , O(\mathbf{s}_B) 
\end{pmatrix}^T
  \in\mathbb{R}^{B\times N_{\mathrm{p}}} , \\
  \boldsymbol{\epsilon} & = &
\begin{pmatrix} 
\boldsymbol{\epsilon}(\mathbf{s}_1) , \boldsymbol{\epsilon}(\mathbf{s}_2) , \cdots, \boldsymbol{\epsilon}(\mathbf{s}_B)
\end{pmatrix}^T
 \in\mathbb{R}^B, 
\end{eqnarray}
where $N_p$ is the total number of parameters, which is about $2\times10^5$ used in the CNN-MPS calculations 

The minimization admits a closed-form solution
\begin{eqnarray}
    \mathrm{d}\theta
&= &\mathrm{diag}(V_{k-1})^{-1/2}\mathbf{O}^{T}S^{-1}(\boldsymbol{\epsilon}-\mathbf{O}\phi_{k-1})
\nonumber \\ 
&& + \phi_{k-1}, \\
S&=&\mathbf{O}\mathrm{diag}(V_{k-1})^{-1/2}\mathbf{O}^T+\lambda I.
\end{eqnarray}
This update scheme can be viewed as a momentum-enhanced and stabilized version of SR.

\subsection{Symmetry implementation}

Imposing symmetries such as lattice reflections and $\pi/2$ rotations provides a systematic route to improving both the expressiveness and the stability of the variational ansatz. In this work, we enforce the lattice point-group symmetry $C_{4v}$ by symmetrizing the wave function as
\begin{equation}
    \Psi^{\mathrm{symm}}(\mathbf{s})
    = \sum_{g \in C_{4v}}\Psi(g\mathbf{s}).
\end{equation}
 This symmetrization improves the accuracy while preserving the internal flexibility of the variational wave function .

\subsection{Marshall sign transformation}

For the antiferromagnetic Heisenberg model
\begin{equation}
  H = J_1 \sum_{\langle ij \rangle }  \vec{S}_i \cdot \vec{S}_j
\end{equation}
on a bipartite lattice, such as the square lattice, the ground-state wave function satisfies the Marshall sign rule~\cite{Antiferromagnetism1955}. If the lattice is partitioned into two sublattices $A$ and $B$, the Marshall sign rule states that the sign of the ground-state wave function is determined by the parity of the number of down spins on one sublattice (e.g., the $A$ sublattice): it is positive for even parity and negative for odd parity. This sign structure can be removed by performing a $\pi$ rotation of all spins on the $A$ sublattice about the $z$ axis. Under this transformation the Hamiltonian becomes
\begin{equation}
  H = J_1 \sum_{\langle ij \rangle }
  \left( -S_i^x S_j^x - S_i^y S_j^y + S_i^z S_j^z \right),
\end{equation}
and the corresponding ground-state wave function should have a uniform sign according to the Marshall sign rule.

Applying this Marshall sign transformation is useful in practice, as it accelerates convergence and improves the stability of the variational optimization. In the presence of frustrating next-nearest-neighbor interactions $J_2$ terms (which remain unchanged under the above transformation), the Marshall sign rule is no longer exact. Nevertheless, the transformation still provides a favorable sign structure for the wave function and remains beneficial for optimization~\cite{PhysRevB.100.125124}. In our simulations, we therefore perform the calculations using the Hamiltonian after this Marshall sign transformation.

\subsection{Monte Carlo samplings} 

 In the Monte Carlo sampling procedure, we generate Markov-chain updates by randomly selecting a bond and proposing a spin exchange when the two spins are in opposite states. Both nearest-neighbor and next-nearest-neighbor bonds are included to enhance the exploration of configuration space. The move is accepted or rejected according to the Metropolis criterion based on the ratio of the wave-function amplitudes.

\begin{figure}
    \centering
    \includegraphics[width=0.9\linewidth]{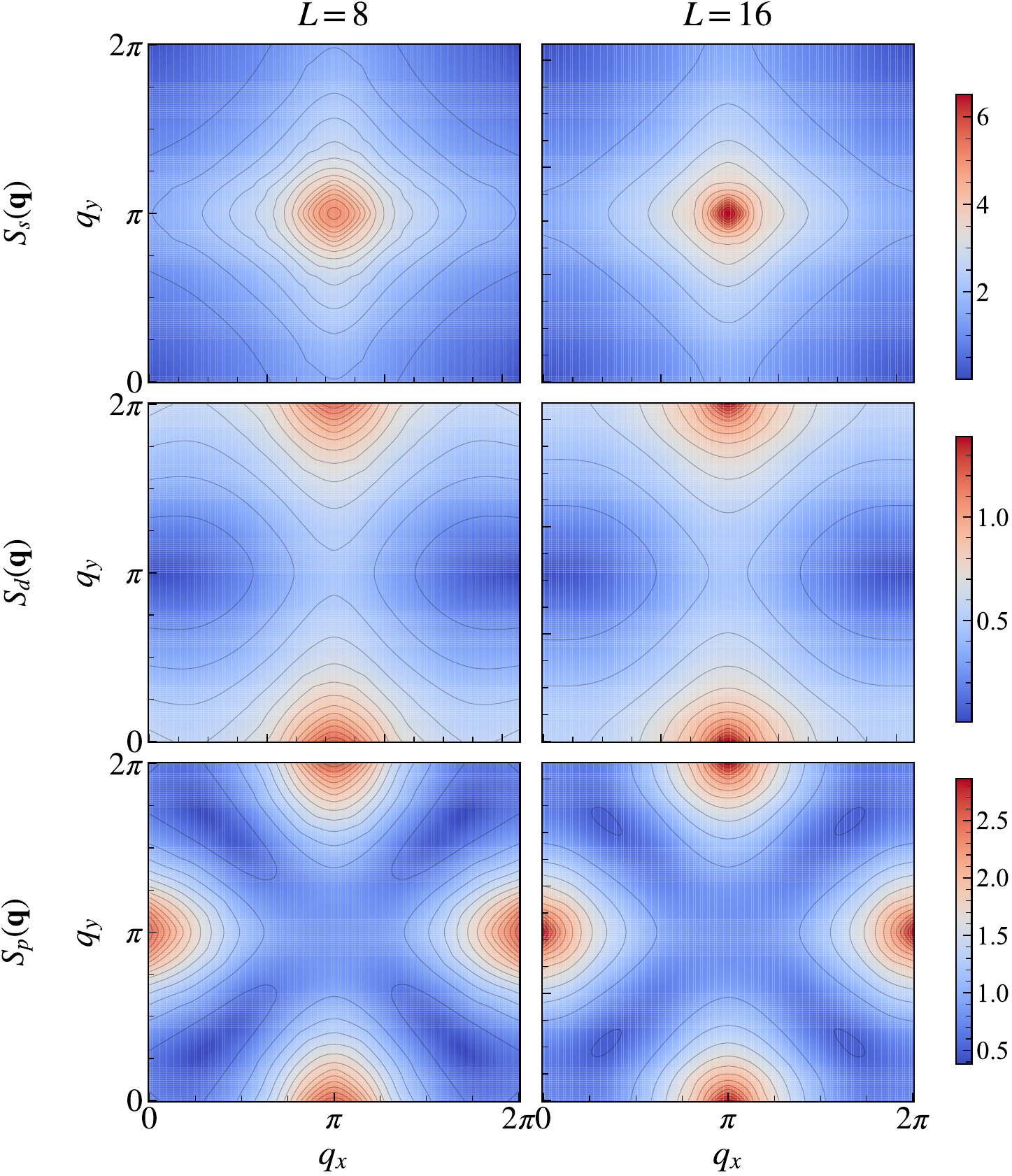}
    \caption{Structure factors of the spin--spin, dimer--dimer, and plaquette--plaquette correlations for the $J_1-J_2$ Heisenberg model on the $L\times L$ square lattice at $J_2/J_1= 0.5$, obtained using CNN-MPS. The dominant momenta occur at $\mathbf{q}=(\pi,\pi)$, $(\pi,0)$, and $(\pi,0)$, as well as at the corresponding symmetry-related points, for the three structure factors, respectively.}
    \label{fig:peak}
\end{figure}

\section{Structure factors}

 Figure~\ref{fig:peak} shows the momentum-space density plots of the spin–spin, dimer–dimer, and plaquette–plaquette structure factors for the $J_1$–$J_2$ Heisenberg model at $J_2/J_1=0.5$, obtained using the CNN-MPS ansatz. The spin structure factor $S_s(\mathbf{q})$ reaches its maximum at $(\pi,\pi)$, resulting from Néel-type antiferromagnetic correlations. The plaquette structure factor $S_p(\mathbf{q})$ shows pronounced peaks at $(\pi,0)$ and $(0,\pi)$, while no peak appears at $(\pi,\pi)$, indicating the absence of plaquette valence-bond (PVB) order. By contrast, the dimer structure factor $S_d(\mathbf{q})$ exhibits clear maxima at $(\pi,0)$, revealing columnar valence-bond correlations in the system.

\bibliography{refs}